\documentclass[a4paper]{article}

\usepackage{INTERSPEECH2022}

\title{Paper Template for INTERSPEECH 2022}
\usepackage{amsthm,amsmath,amssymb}
\usepackage{multirow}
\usepackage{enumerate}
\usepackage{makecell}
\usepackage{cite}

\usepackage{mathrsfs}
\usepackage{float}  
\usepackage{subfigure}  
\usepackage{epsfig}
\usepackage{verbatim}
\usepackage{bigstrut}

\title{The HCCL System for the NIST SRE21}
\name{Zhuo Li{\rm\textsuperscript{*}}\thanks{\textsuperscript{*}equal contribution}, Runqiu Xiao{\rm\textsuperscript{*}}, Hangting Chen, Zhenduo Zhao, Zihan Zhang, Wenchao Wang}
\address{
 Key Laboratory of Speech Acoustics and Content Understanding, \\ Institute of Acoustics, Chinese Academy of Sciences, Beijing, China\\
 University of Chinese Academy of Sciences, Beijing, China 
}
\email{\{lizhuo,xiaorunqiu\}@hccl.ioa.ac.cn}

\begin{document}

\maketitle
\begin{abstract}
This paper describes the systems developed by the HCCL team for the NIST 2021 speaker recognition evaluation (NIST SRE21).
We first explore various state-of-the-art speaker embedding extractors combined with a novel circle loss to obtain discriminative deep speaker embeddings.
Considering that cross-channel and cross-linguistic speaker recognition are the key challenges of SRE21, we 
introduce several techniques to reduce the cross-domain mismatch.
Specifically, Codec and speech enhancement are directly applied to the raw speech
to eliminate the codecs and the environment noise mismatch.
We denote these methods that work directly on raw audio to eliminate the relatively explicit mismatch collectively as data adaptation methods. Experiments show that data adaption methods achieve 15\% improvements over our baseline. 
Furthermore, some popular back-ends domain adaptation algorithms are deployed on speaker embeddings to  alleviate speaker performance degradation caused by the implicit mismatch.
Score calibration is a major failure for us in SRE21. The reason is that score calibration with excessive  parameters easily leads to overfitting.

\end{abstract}

\section{Introduction}
Speaker recognition aims to identify or verify an individual’s identity from samples of his/her voice, which has developed rapidly over the years. 
The series of Speech Recognition Evaluations (SRE) conducted by the National Institute of Standards and Technology (NIST) usually reflects the most advanced technology in the field of speaker recognition.
The 2021 Speaker Recognition Evaluation (SRE21)\cite{sadjadi2021nist} is the latest of the series of speaker recognition evaluations. Unlike the previous SREs, SRE21 brought serious cross-linguistic and cross-channel problems between enrollment and test, althought it still focus on cross-domain problems.

In this paper, we describe our systems developed for the NIST SRE21.
All subsystems use a typical two-stage pipeline consisting of a strong discriminative speaker embedding extractor and a domain-specific PLDA backend\cite{plda,plda02}. 
In addition to the typical Extended TDNN\cite{xvectorformal,xvectorbefore}, ECAPA\cite{ecapa}, and SE-ResNet\cite{resnet} models, we introduce a powerful network topology, RepVGG and a more discriminative margin-based angular loss function, which is proposed in \cite{ding2021repvgg}.
To improve the robustness of extractors, we increase the amount and diversity of training data through data augmentation.
Cross-domain remains one major challenge for speaker recognition. Some cross-domain situations are obvious, observable or easily changed, e.g., sample rate, duration, noise. While others are latent, difficult to observe or change, e.g., channel, language, recording conditions. For the latter, regularization and adaptation working on embeddings or back-ends to eliminate the mismatch are the best approach\cite{lee2019coral+}. For the former, however, it is perhaps better to eliminate the mismatch by working directly on audio, that is, data adaptation. In this paper, we explore ways to eliminate mismatches directly on audio from two perspectives, one is codec, the other is denoise. The complementarity of data adaptation and back-ends adaptation algorithms is explored afterwards.
Finally, score calibration and fusion are described and analyzed. By comparing with other methods, we analyze the reasons for failure.

The rest of this paper is organized as follows: data usage and setting are briefly introduced in Section 2, Section 3 describes deep speaker embedding extractors in detail. Data adaptation methods are introduced in Section 4, followed by back-ends domain adaptation algorithms in Section 5. Next, calibration and fusion are described in Section 6. Finally, section 7 conclude the paper.

\section{Data \& Settings }
\label{sec:format}
\subsection{Training data}
The training data for SRE21 consist of the NIST SRE CTS Superset\cite{cts}, SRE16 evaluation set\cite{sadjadi20172016} and the VoxCeleb corpus\cite{vox1,vox2}.
The NIST speaker recognition evaluation (SRE) conversational telephone speech (CTS) superset 
is extracted from prior SRE datasets(SRE1996-2012), which contains 6867 speakers. Each segment contains approximately 10 to 60 seconds of speech and most of the speech segments are spoken in English.
NIST SRE16 set provides two sets of development data and one unlabeled set. The dev set consists of 20 individuals, ten of whom speak Mandarin and ten of whom speak Cebuano. Similar to the dev set, the eval set contains 201 individuals, 100 speak Tagalog and 101 speak Cantonese.
VoxCeleb 1\&2 are large scale speaker recognition datasets collected from YouTube. VoxCeleb 1 includes dev and test parts, with 1211 and 40 speakers separately. While VoxCeleb2 contains 5994 speakers in dev set and 120 speakers in test set.
Most of the utterance lengths of VoxCeleb distributes are in the [4s,15s] range.
Due to its short duration, we concatenate the subsegments belonging to the same original video into a unique segment to balance the duration of each segment during the training stage, denoted as VoxCat.

\subsection{Dev and eval set}
The SRE21 development and evaluation set consist of conversational telephone speech and audio from video (AfV). Similar to the recent SREs(SRE18,SRE19,SRE20), differences in terms of languages and channels cause a considerable mismatch between the trained model and test data. However, in SRE21, the mismatch between enroll and test audio in terms of languages and channels is more challenging, especially for target trials.
After analyzing the audio of the dev and eval sets, we briefly summarize the main difficulties: i) language mismatch between the training set and test set. ii) channel mismatch between the training set and test set. iii) language mismatch between enrollment and test. iv)channel mismatch between enrollment and test.

\subsection{Augmentation data}
In order to increase the diversity of the acoustic conditions in the training set, we create four corrupted copies of the original recordings, which are corrupted by either digitally adding noise (i.e., babble, general noise, music) or convolving with simulated and measured room impulse responses (RIRS). In addition, to mitigate the difference in domains between VoxCat and NIST SRE sets, GSM AMR codec at 6.7 and 4.75 kps augmentation are added to VoxCat.

\subsection{Experiments setting.}
\textbf{Data preprocessing} An energy-based and harmonics-based SAD is used to drop the non-speech frames. 
For speech parameterization, we extract 81-dimensional Fbank and 80-dimensional MFCC from 25ms frames every 10ms using 80-channel filterfbank spanning the frequency range 40-3800 Hz and 20-3700Hz, respectively. In addition, 3-dimension of the pitch features are concatenated. Finally, We applied cepstral mean-normalization (CMN) with a sliding window of up to 3 seconds on these acoustic features.

\textbf{Training strategy.}
The mini-batch size is 64. Stochastic gradient descent (SGD) with momentum
0.9 is utilized. The learning rate is set to 0.1, 0.01, 0.001 and is switched when the training loss plateaus. The chunk-size $L$ is randomly sampled from the interval
[$L1$;$L2$], and the interval is set to [400,800], [600,1000] and [800,1000/1200] in the three training stages. 

\section{Extractors}

\subsection{Encoder network}
\subsubsection{ETDNN}
The Extended Time Delay Neural Network architecture (E-TDNN) has been introduced in \cite{xvectorbefore,xvectorformal} and widely used in the past years. Similar to the TDNN x-vector network, E-TDNN is composed of a sequence of time-delay layers and a pooling layer. Meanwhile, E-TDNN slightly enlarges  temporal context and interleaves fully-connected layers between the TDNN layers. This architecture has been found to perform greatly in SRE18 and SRE19 benchmarks. In our systems, the number of channels and embedding size are set to 1024, an attentive statistics pooling layer\cite{okabe2018attentive} and two fully-connected layers transfer frame-level features to segment-level features. The nonlinearities are Leaky rectified linear units (leaky ReLUs).

%
%

\subsubsection{ECAPA}
Emphasized Channel Attention, Propagation and Aggregation in TDNN (ECAPA-TDNN), newly proposed in \cite{ecapa}, is an enhanced version of the standard TDNN architecture. It integrates a Res2Net module to  
enhance the central convolutional layer so that it can process multi-scale features by constructing hierarchical residual-like connections within. The 1-dimensional SE-block and multi-layer feature aggregation are integrated to extract complementary information from the shallow feature maps.
In our systems, Res2Net module\cite{gao2019res2net} is replaced by Hierarchical-Split (HS) module\cite{hsresnet}. 
Through our theoretical derivation\cite{9688119} and experimental validation, it is found that little difference exists between stacked Res2Net modules and stacked HS modules, either in terms of expressive capacity or experimental results. While, HS modules performs slightly better on minDCF.
We use our implementation of ECAPA-TDNN with the following parameters: the number of HS-ResNet blocks are set to 4 with dilation values 2,3,4,5 for blocks; the number of channels and embedding size are set to 1024.

\subsubsection{ResNet-SE}
As one of the most classical CNN network, ResNet with squeeze and excitation (SE) block\cite{hu2018squeeze} has achieved great performance in speaker recognition. In this system, we adopt ResNet34-SE with ASP layer and the channel configurations of residual block are (32,64,128,256) and (64,128,256,512).

\begin{table*}[htbp]
  \centering
  \caption{Results of systems in SRE21 sets}
\setlength{\tabcolsep}{1.5mm}{
    \begin{tabular}{ccc|cccc|cccc|cccc}
    \toprule
          & \multirow{3}[4]{*}{System} & \multirow{3}[4]{*}{Feature} & \multicolumn{4}{c|}{PLDA} & \multicolumn{4}{c|}{ +data\_adaptation} & \multicolumn{4}{c}{ +CORAL+} \\
\cmidrule{4-15}          &       &       & \multicolumn{2}{c}{dev} & \multicolumn{2}{c|}{eval} & \multicolumn{2}{c}{dev} & \multicolumn{2}{c|}{eval} & \multicolumn{2}{c}{dev} & \multicolumn{2}{c}{eval} \\
          &       &       & EER   & min$C_p$ & EER   & min$C_p$ & EER   & min$C_p$ & EER   & min$C_p$ & EER   & min$C_p$ & EER   & min$C_p$ \\
    \midrule
    1     & RepVgg & Fbank & 7.43  & 0.517 & 7.37  & 0.544 & 5.71  & \textbf{0.349}  & 5.75  & 0.414 & 4.14  & \textbf{0.272}  & 4.43  & \textbf{0.315} \\
    2     & ResNet34-32 & Fbank & 7.34  & 0.621 & 6.95  & 0.543 & 5.67  & 0.470  & 5.55  & 0.426 & 4.08  & 0.318 & 4.08  & 0.320 \\
    3     & ETDNN & Fbank & 6.68  & 0.465 & 7.21  & 0.508 & 5.94  & 0.381 & 6.46  & 0.454 & 4.63  & 0.282 & 5.26  & 0.396 \\
    4     & ETDNN & MFCC  & 7.06  & \textbf{0.439} & 7.54     & 0.527     & 6.67  & 0.415 & 7.05  & 0.515 & 5.00     & 0.311 & 5.66  & 0.432 \\
    5     & ECAPA & Fbank & 8.74  & 0.720  & 9.69  & 0.739 & 6.22  & 0.520  & 6.93  & 0.559 & 4.25  & 0.353 & 4.76  & 0.391 \\
    6     & ResNet34-64 & Fbank & 6.75  & 0.566 & 6.19  & \textbf{0.463} & 5.89  & 0.430  & 5.49  & \textbf{0.402} & 4.72  & 0.325  & 4.10   & 0.328 \\
    \bottomrule
    \end{tabular}}%
  \label{t1}%
\vspace{-5mm}
\end{table*}%

\subsubsection{RepVGG}
RepVGG are proposed in \cite{ding2021repvgg} and achieves great performance in computer vision field. It decouples a training-time multibranch topology with an inference-time plain architecture by using structural re-parameterization, which replaces parallel convolution modules with one convolution kernel. Our prior work in SRE20 challenge and \cite{zhao2021speakin} have proven its power in speaker verification and recognition.

\begin{figure}[!htbp]
\centering
\setlength{\belowcaptionskip}{0pt}
\vspace{-2mm}
\includegraphics[width=0.9\linewidth]{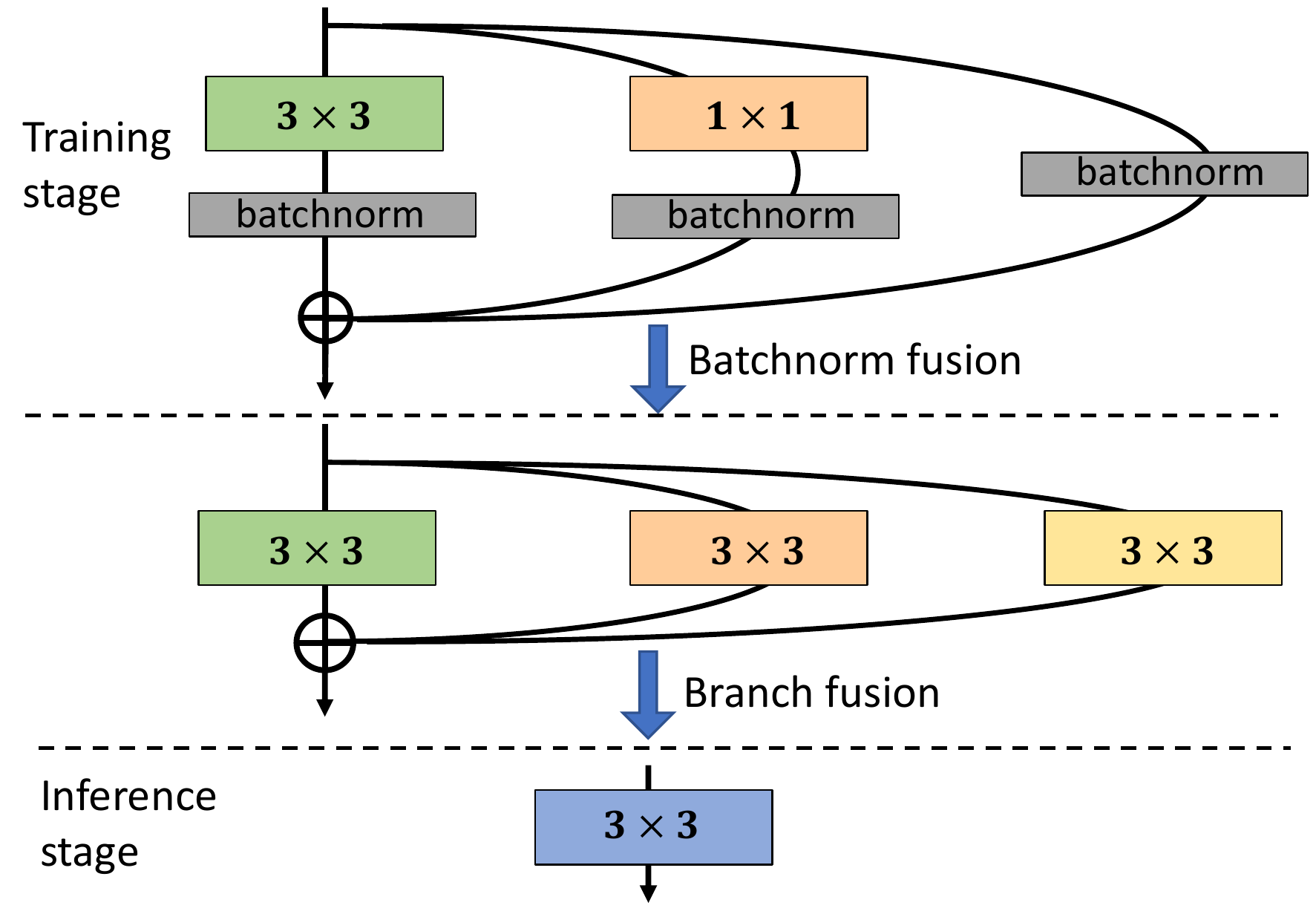}
\vspace{-3mm}
\caption{Structural re-parameterization of a RepVGG
block.}
\label{repvgg}
\vspace{-3mm}
\end{figure}

As Fig.~\ref{repvgg} shows, in the training stage, a multi-branch topology, named as RepVGG block, is adopt with multi-scale convolutions, $1\times1$ and $3\times3$ convolutions and identity map. These operations with various receptive field can enrich the feature space and enhance the representational capacity of models. Then, in the inference stage, RepVGG block with multi-branch is equivalently transformed into a single conv to accelerate inference.

Then, we describe how to convert a multi-branch RepVGG block into a single $3\times3$ conv layer.
\vspace{-2mm}
\begin{equation}
\begin{aligned}
&M^{out} = \sum_{k=0,1,3}bn(M^{in}\otimes W^{(k)},\mu^{(k)},\sigma^{(k)},\gamma^{(k)},\beta^{(k)})\\
&bn(M,\mu,\sigma,\gamma,\beta)_{:,j,:,:} = (M_{:,j,:,:} - \mu_j)\gamma_j/\sigma_j + \beta_j \\
\end{aligned}
\end{equation}
\vspace{-2mm}
\begin{equation}
\begin{aligned}
\bm{Let}:\ & W^{'}_{j,:,:,:}=\sum_{k=0,1,3}\gamma^{(k)}_j/\sigma^{(k)}_jW_{j,:,:,:}, \\
&b^{'}_{j,:,:,:}=\sum_{k=0,1,3}-\mu^{(k)}_i\gamma^{(k)}_j/\sigma^{(k)}_j+\beta^{(k)}_j\\
\bm{Then}:\ & M^{out}=M^{in}\otimes W^{'}+b^{'}
\end{aligned}
\label{eqrep}
\end{equation}
Formally, we use $W^{(k)} \in R^{C_1\times C_2\times K\times K}$ to denote the kernel of a $K\times K$ conv layer with $C_1$ input channels and $C_2$ output channels, $M^{in} \in R^{N\times C_1\times H_1\times W_1}$, $M^{out} \in R^{N\times C_2\times H_2\times W_2}$ be the input and output, respectively, and $\otimes$ be the convolution operator. 
Let $j$ be the channel index, $\mu_j$, $\sigma_j$, $\gamma_j$, $\beta_j$ be the accumulated mean, standard deviation and learned scaling factor and bias of the BN layer, respectively.

After Eq.~\ref{eqrep}, the multi-branch block is converted into a single block. In our system, RepVGG-B1 with 64 channels is used.

\subsection{Loss Function}
To enhance the intra-class compactness of speaker embeddings, we use circle loss instead of traditional softmax loss or angular softmax loss. Circle loss was firstly proposed in ~\cite{sun2020circle} and then used in speaker recognition in our paper~\cite{xiao21b_interspeech}. Compared with the previous angular loss like Additive Margin Softmax(Am-Softmax), circle loss has flexible optimization and definite convergence status. 
The proposed circle loss is defined as:

\vspace{-2.5mm}
\begin{equation}
\centering
\begin{aligned}
L_{\rm circle}=-{\rm log}\frac{e^{s\cdot (m^2-(1-cos\theta_{y_i})^2)}}{e^{s\cdot (m^2-(1-cos\theta_{y_i})^2)} + \sum_{j\ne i}e^{s\cdot ((cos\theta_j)^2-m^2)}}
\end{aligned}
\label{eq1}
\end{equation}
The $\theta_{y_i}$ is the angle between the input vector and the weight vector of class $y_i$ among the $C$ target classes. We set $s=60$, $m=0.35$ for CNN system and $s=60$, $m=0.40$ for TDNN. 

\section{Data adaptation}
\label{sec:pagestyle}
As stated before, some cross-domain situations are observable and easily changed, while others are latent and difficult to change. For the former, it is perhaps better to eliminate by directly working on raw audio. We notice that the AFV segments, which are stored as 16 bit FLAC files, are different from the CTS segments, which are stored as 8 bit a-law SPH files.
These differences are observable in encoding and background noise. 


\subsection{Codec}
It is noticed that the AFV segments are Free Lossless Audio Codec, while, the CTS segments are alaw. There will exist large encoding mismatch if they are converted directly to 8kHz-16bit-PCM.
Instead of downsampling directly the flac-formatted speech into 8k-16bit wav-formatted speech, we first convert it into 8k-alaw-formatted speech and then into 8k-16bit wav-formatted speech.
By analyzing the cosine similarity between the same speech with different transcoding ways on different systems, it is found that nearly half of the speech similarities are below 0.95, with a minimum of 0.8.
It indicates that codec influences the speaker embeddings.
\begin{table}[htbp]
  \centering
  \caption{Results of data adaptation on RepVGG model}
    \begin{tabular}{c|cc|cc}
    \toprule
    \multirow{3}[4]{*}{} & \multicolumn{4}{c}{dev} \\
\cmidrule{2-5}          & \multicolumn{2}{c|}{Cos} & \multicolumn{2}{c}{PLDA} \\
          & \textit{EER} & min$C_p$ & \textit{EER} & min$C_p$ \\
    \midrule
    ini   & 8.48  & 0.506 & 7.43 & 0.517 \\
    \midrule
    denoise & 8.20   & 0.506 & 7.57 & 0.474 \\
    \midrule
    codec & 8.40   & 0.485 & 6.17 & 0.419 \\
    \midrule
    \multicolumn{1}{p{7.19em}|}{denoise$_{clean}$+codec} & 7.31  & 0.463 & 5.51 & 0.365 \\
    \midrule
    denoise$_{cts}$+codec & 7.11  & \textbf{0.459} & 5.71 & \textbf{0.349} \\
    \bottomrule
    \end{tabular}%
  \label{t2}%
\vspace{-4mm}
\end{table}%

\begin{table*}[!th]
\vspace{-4mm}
  \centering
  \caption{Results of systems after different partition calibrations}
\setlength{\tabcolsep}{1.4mm}{
    \begin{tabular}{cc|ccc|ccc|ccc|ccc}
    \toprule
    \multicolumn{2}{c|}{\multirow{3}[2]{*}{System}} & \multicolumn{6}{c|}{partition calibration (channel and gender)}    & \multicolumn{6}{c}{ partition calibration (only channel)} \\
    \multicolumn{2}{c|}{} & \multicolumn{3}{c|}{dev} & \multicolumn{3}{c|}{eval} & \multicolumn{3}{c|}{dev} & \multicolumn{3}{c}{eval} \\
    \multicolumn{2}{c|}{} & EER   & $\min C_p$ &  $act C_p$   & EER   & $\min C_p$  &  $act C_p$   & EER   & $\min C_p$ & $act C_p$   & EER  & $\min C_p$ & $act C_p$ \\
    \midrule
    1     & RepVgg & 3.30   & \textbf{0.223 } & 0.227  & 3.57  & \textbf{0.282} & 0.316 & 3.30   & 0.202  & 0.207  & 3.87  & \textbf{0.267} & 0.317 \\
    2     & ResNet34-32 & 3.62  & \textbf{0.237} & 0.241 & 3.87  & \textbf{0.294} & 0.307 & 3.48  & 0.229 & 0.236 & 3.57  & \textbf{0.278} & 0.294 \\
    3     & ETDNN & 3.35  & 0.245 & 0.256 & 4.51  & 0.355 & 0.373 & 3.79  & 0.252 & 0.267 & 4.78  & 0.359 & 0.414 \\
    4     & ETDNN(M) & 3.74  & 0.280  & 0.282  & 5.28  & 0.415 & 0.442 & 4.23  & 0.275 & 0.279 & 5.14  & 0.390  & 0.440 \\
    5     & ECAPA & 3.81  & 0.269  & 0.276  & 4.00     & 0.341 & 0.414 & 3.50   & 0.243 & 0.249 & 3.80   & \textbf{0.288} & 0.304 \\
    6     & ResNet34-64 & 3.93  & 0.255  & 0.263  & 3.74  & 0.326 & 0.402 & 3.90   & 0.235  & 0.238  & 3.75  & 0.304 & 0.358 \\
    \midrule
    \multicolumn{2}{c|}{fusion} & 2.58  & 0.186  & 0.189  & 2.90   & 0.341 & 0.373 & 2.69  & 0.175 & 0.182 & \textbf{3.00} & \textbf{0.228} & \textbf{0.231} \\
    \midrule
    \multicolumn{2}{c|}{\textbf{failed-calibration(submit)}} & 2.27  & 0.168  & 0.170  & \textbf{3.05} & \textbf{0.398} & \textbf{0.560} & —     & —     & —     & —     & —     & — \\
    \bottomrule
    \end{tabular}}%
  \label{t3}%
\vspace{-2mm}
\end{table*}%

\subsection{Denoise}
After analyzing the AFV segments, we notice that many flac-files contain strong low frequency narrow band noise and background noise, which causes a huge mismatch between cts-files.
We deploy a hybrid denoise model, which contains convolutional-deconvolutional U-net structure~\cite{hu2020dccrn} and dual-path long short-term memory (DP-RNN~\cite{luo2020dual}) to eliminate the background noise. As suggested in ~\cite{li2021icassp}, the complex domain method was found to achieve better mean opinion scores. In the hyrid model, the input audio is first transformed into time-frequency spectrum using short-time frequency transform. The spectrum is convolved by 2 sequential convolutional layers into a 3-D (channel, frequency and time) high-level representation. The DP-RNN is built using 2 stacked dual-path blocks, each of which consists of a bidirectional LSTM on frequency dimension, a global layer normalization, a unidirectional LSTM on time dimension and a global layer normalization. After processed by the DP-RNN module, the 3-D feature map is fed into 2 deconvolutional layers to generate complex masks~\cite{hu2020dccrn}. The enhanced signal can be obtained by inverse STFT operated on the enhanced spectrum.
\begin{figure}[!htbp]
\centering
\setlength{\belowcaptionskip}{0pt}
\includegraphics[width=1.0\linewidth]{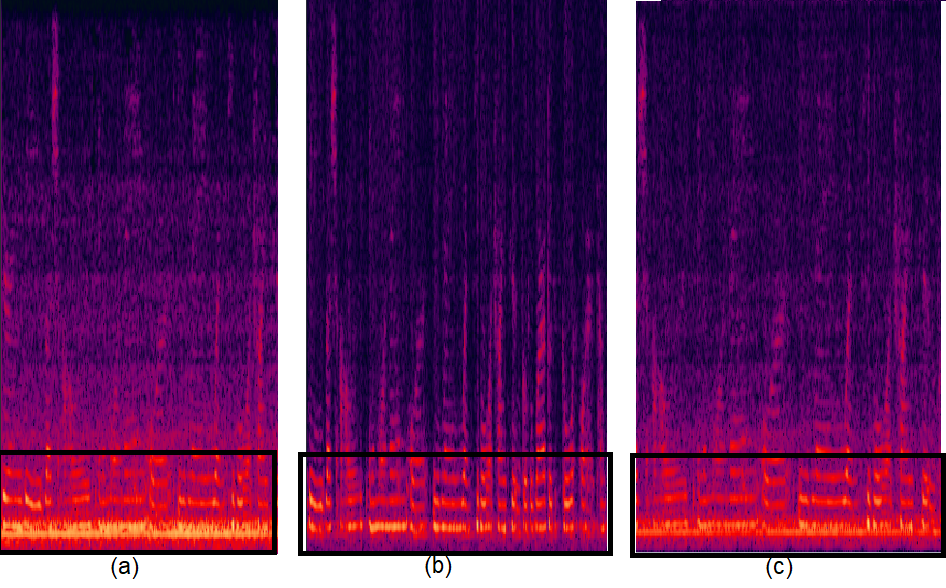}
\vspace{-5mm}
\caption{The speech spectrum before and after denoising. (a) is the initial spectrum, (b) is the spectrum after denoise by model trained with clean data, (c) is the spectrum after denoise by model trained with CTS data.}
\label{de}
\vspace{-2mm}
\end{figure}

Fig.~\ref{de} shows the speech spectrum before and after denoising. We firstly use the model trained with clean data(e.g., AIShell~\cite{8384449,du2018aishell}). The speech after denoising is better than initial wav, either on subjective metric of speech enhancement or objective metric of speaker recognition, as shown in Fig.~\ref{de}(b).
Due to the fixed training condition, we use CTS-data to train models. As Fig.~\ref{de}(b) shows,  the model trained with CTS-data is worse than the model trained with our-data, either on subjective metric or objective metric of speech enhacement.
However, its performance on SV is similar, results are present in Table~\ref{t2}.

\subsection{Results and analysis}
Results of data adaptation on RepVGG model are shown in Table~\ref{t2}, and results of all systems are shown in Table~\ref{t1}. Three things are worth noting. (i) Different data domain adaptation methods are complementary to each other,
(ii) Data adaptation achieve little improvements on Cos, about 10\% in $\min C_p$, while large improvements on PLDA, about 30\% in $\min C_p$ in total.
(iii) Data adaptation works on all models.

\section{Back-ends domain adaption}

The CORAL algorithm is proposed to align the covariance between different domain embeddings via a whitening and re-coloring process. In \cite{lee2019coral+}, the author integrate CORAL into PLDA to adapt the between-class and within-class covariance matrices of PLDA model. 
\begin{equation}
\centering
\begin{aligned}
&\Phi_{out} = \Phi_{w,out} + \Phi_{b,out} \quad
\Phi_{in} = \Phi_{w,in} + \Phi_{b,in} \\
&\quad\quad\quad\quad\quad\quad\quad M = \Phi_{in}^{1/2}\Phi_{out}^{-1/2} \\
&\Phi_{w,c} = M^T\Phi_{w,out}M \quad
\Phi_{b,c} = M^T\Phi_{b,out}M \\
\end{aligned}
\end{equation}
Then, to improve the robustness of adaptation, 
\cite{lee2019coral+} introduce additional adaptation parameter and regularization to the adaptation equation, named as CORAL+.
For convenience, the training set is referred to as out-of-domain (OOD) since it differs from the test set, which is the area of interest and hence referred to as in-domain (InD).
\begin{equation}
\centering
\begin{aligned}
&B_{*}^T\Phi_{*,o}B_{*} = I \quad B_{*}^T\Phi_{*,c}B_{*} = E_{*}
\, Here:\,  *\in \{b,w\}\\
&\Phi_{b,c+} = \Phi_{b,out}\ +\  \gamma B_b^{-T}max\{E_b,I\}B_b^{-1} \\
&\Phi_{w,c+} = \Phi_{w,out}\ +\  \beta B_w^{-T}max\{E_w,I\}B_w^{-1} \\
\end{aligned}
\end{equation}

Here, $\Phi_{out}$ and $\Phi_{in}$ are the covariance matrices of the OOD and InD data, respectively. $M$ is the transformation matrix. 
$B$ is an orthogonal matrix to diagonalise $\Phi_{*,o}$ and $\Phi_{*,c}$, as equation shows. Finally,$\Phi_{b,c+}$ and $\Phi_{w,c+}$ are the between- and within-class covariance matrices after the CORAL+ adaptation. $\{\gamma,\beta\}$ are the adaptation parameters.

Results of all systems are shown in Table~\ref{t1}. It is found that CORAL+ achieve about 10\%-25\% improvements in $\min C_p$.

\section{Calibration and Fusion}
Because the distribution of scores is different in different parts, we use partition calibration. We calibrate the scores in different test conditions separately by BOSARIS Toolkit~\cite{brummer2013bosaris} after we obtain pseudo-labels of language and gender. Besides, the scores are calibrated with various duration characteristics, which include the duration of segment, the l2 norm of embeddings~\cite{thienpondt2021idlab}, named as Quality Measurements(QM).

However, these leads to a significant deviation between the performance of dev and eval set, either in $\min C_p$ or $act C_p$. The results are not present in Table~\ref{t3}.
The major reason is that we do not set constraints and give too many parameters leading to overfitting on the dev set.
Thus, we use channel and gender information to calibrate scores after removing QMs and language pseudo-labels. Results are shown in \textit{partition calibration (channel and gender)} of Table~\ref{t3}. It is obvious that a deviation still exists between $\min C_p$ and $act C_p$, which should be equal in theory if calibration is good.
After analyzing scores, we find that not enough male trials in dev set are the major reason. After removing gender information from calibration, the deviation between $\min C_p$ or $act C_p$ in the eval set reduces significantly. Results are shown in \textit{partition calibration (only channel)} part of Table~\ref{t3}.
Thus, we can draw two conclusions, (i) simpler, more robust, (ii) the number of trials in dev set is important.

\section{Conclusions}

This paper presents systems developed by HCCL team for the NIST SRE21. We explore various strong speaker embedding extractors and circle loss to get discrimitive embeddings. Data adaptation methods are proposed to mitigate the cross-domain mismatch. 
Results show that observable and easily changed mismatches such as noise and codecs can be mitigated directly from raw audio. 
Also, back-ends adaptation methods is necessary in cross-domain situations. In score calibration, too many parameters leads to overfitting on dev set, and the number of trials in dev set affects the robustness of score calibration. Finally, score fusion of different structure models achieve further improvements.
%

\bibliographystyle{IEEEtran}

\bibliography{mybib}

\begin{thebibliography}{10}
\providecommand{\url}[1]{#1}
\csname url@samestyle\endcsname
\providecommand{\newblock}{\relax}
\providecommand{\bibinfo}[2]{#2}
\providecommand{\BIBentrySTDinterwordspacing}{\spaceskip=0pt\relax}
\providecommand{\BIBentryALTinterwordstretchfactor}{4}
\providecommand{\BIBentryALTinterwordspacing}{\spaceskip=\fontdimen2\font plus
\BIBentryALTinterwordstretchfactor\fontdimen3\font minus
  \fontdimen4\font\relax}
\providecommand{\BIBforeignlanguage}[2]{{%
\expandafter\ifx\csname l@#1\endcsname\relax
\typeout{** WARNING: IEEEtran.bst: No hyphenation pattern has been}%
\typeout{** loaded for the language `#1'. Using the pattern for}%
\typeout{** the default language instead.}%
\else
\language=\csname l@#1\endcsname
\fi
#2}}
\providecommand{\BIBdecl}{\relax}
\BIBdecl

\bibitem{sadjadi2021nist}
O.~Sadjadi, C.~Greenberg, E.~Singer, L.~Mason, and D.~Reynolds, ``Nist 2021
  speaker recognition evaluation plan,'' 2021.

\bibitem{plda}
S.~Ioffe, ``Probabilistic linear discriminant analysis,'' in \emph{European
  Conference on Computer Vision}.\hskip 1em plus 0.5em minus 0.4em\relax
  Springer, 2006, pp. 531--542.

\bibitem{plda02}
P.~Kenny, ``Bayesian speaker verification with, heavy tailed priors,''
  \emph{Proc. Odyssey 2010}, 2010.

\bibitem{xvectorformal}
D.~Snyder, D.~Garcia-Romero, G.~Sell, D.~Povey, and S.~Khudanpur, ``X-vectors:
  Robust dnn embeddings for speaker recognition,'' in \emph{2018 IEEE
  International Conference on Acoustics, Speech and Signal Processing
  (ICASSP)}.\hskip 1em plus 0.5em minus 0.4em\relax IEEE, 2018, pp. 5329--5333.

\bibitem{xvectorbefore}
D.~Snyder, D.~Garcia-Romero, D.~Povey, and S.~Khudanpur, ``Deep neural network
  embeddings for text-independent speaker verification,'' in \emph{Proc.
  Interspeech 2017}, 2017, pp. 999--1003.

\bibitem{ecapa}
B.~Desplanques, J.~Thienpondt, and K.~Demuynck, ``Ecapa-tdnn: Emphasized
  channel attention, propagation and aggregation in tdnn based speaker
  verification,'' \emph{Proc. Interspeech 2020}, pp. 3830--3834, 2020.

\bibitem{resnet}
K.~{He}, X.~{Zhang}, S.~{Ren}, and J.~{Sun}, ``Deep residual learning for image
  recognition,'' in \emph{2016 IEEE Conference on Computer Vision and Pattern
  Recognition (CVPR)}, 2016, pp. 770--778.

\bibitem{ding2021repvgg}
X.~Ding, X.~Zhang, N.~Ma, J.~Han, G.~Ding, and J.~Sun, ``Repvgg: Making
  vgg-style convnets great again,'' in \emph{Proceedings of the IEEE/CVF
  Conference on Computer Vision and Pattern Recognition}, 2021, pp.
  13\,733--13\,742.

\bibitem{lee2019coral+}
K.~A. Lee, Q.~Wang, and T.~Koshinaka, ``The coral+ algorithm for unsupervised
  domain adaptation of plda,'' in \emph{ICASSP 2019-2019 IEEE International
  Conference on Acoustics, Speech and Signal Processing (ICASSP)}.\hskip 1em
  plus 0.5em minus 0.4em\relax IEEE, 2019, pp. 5821--5825.

\bibitem{cts}
S.~O. Sadjadi, ``Nist sre cts superset: A large-scale dataset for telephony
  speaker recognition,'' \emph{arXiv preprint arXiv:2108.07118}, 2021.

\bibitem{sadjadi20172016}
S.~O. Sadjadi, T.~Kheyrkhah, A.~Tong, C.~S. Greenberg, D.~A. Reynolds,
  E.~Singer, L.~P. Mason, J.~Hernandez-Cordero \emph{et~al.}, ``The 2016 nist
  speaker recognition evaluation.'' in \emph{Interspeech}, 2017, pp.
  1353--1357.

\bibitem{vox1}
A.~Nagrani, J.~S. Chung, and A.~Zisserman, ``Voxceleb: a large-scale speaker
  identification dataset,'' \emph{arXiv preprint arXiv:1706.08612}, 2017.

\bibitem{vox2}
J.~S. Chung, A.~Nagrani, and A.~Zisserman, ``{VoxCeleb2: Deep Speaker
  Recognition},'' in \emph{Proc. Interspeech 2018}, 2018, pp. 1086--1090.

\bibitem{okabe2018attentive}
K.~Okabe, T.~Koshinaka, and K.~Shinoda, ``Attentive statistics pooling for deep
  speaker embedding,'' \emph{Proc. Interspeech 2018}, pp. 2252--2256, 2018.

\bibitem{gao2019res2net}
S.~Gao, M.-M. Cheng, K.~Zhao, X.-Y. Zhang, M.-H. Yang, and P.~H. Torr,
  ``Res2net: A new multi-scale backbone architecture,'' \emph{IEEE transactions
  on pattern analysis and machine intelligence}, 2019.

\bibitem{hsresnet}
P.~Yuan, S.~Lin, C.~Cui, Y.~Du, R.~Guo, D.~He, E.~Ding, and S.~Han,
  ``Hs-resnet: Hierarchical-split block on convolutional neural network,''
  \emph{arXiv preprint arXiv:2010.07621}, 2020.

\bibitem{9688119}
Z.~Li, C.~Fang, R.~Xiao, W.~Wang, and Y.~Yan, ``Si-net: Multi-scale
  context-aware convolutional block for speaker verification,'' in \emph{2021
  IEEE Automatic Speech Recognition and Understanding Workshop (ASRU)}, 2021,
  pp. 220--227.

\bibitem{hu2018squeeze}
J.~Hu, L.~Shen, and G.~Sun, ``Squeeze-and-excitation networks,'' in
  \emph{Proceedings of the IEEE conference on computer vision and pattern
  recognition}, 2018, pp. 7132--7141.

\bibitem{zhao2021speakin}
M.~Zhao, Y.~Ma, M.~Liu, and M.~Xu, ``The speakin system for voxceleb speaker
  recognition challange 2021,'' \emph{arXiv preprint arXiv:2109.01989}, 2021.

\bibitem{sun2020circle}
Y.~Sun, C.~Cheng, Y.~Zhang, C.~Zhang, L.~Zheng, Z.~Wang, and Y.~Wei, ``Circle
  loss: A unified perspective of pair similarity optimization,'' in
  \emph{Proceedings of the IEEE/CVF Conference on Computer Vision and Pattern
  Recognition}, 2020, pp. 6398--6407.

\bibitem{xiao21b_interspeech}
R.~Xiao, X.~Miao, W.~Wang, P.~Zhang, B.~Cai, and L.~Luo, ``{Adaptive Margin
  Circle Loss for Speaker Verification},'' in \emph{Proc. Interspeech 2021},
  2021, pp. 4618--4622.

\bibitem{hu2020dccrn}
Y.~Hu, Y.~Liu, S.~Lv, M.~Xing, S.~Zhang, Y.~Fu, J.~Wu, B.~Zhang, and L.~Xie,
  ``Dccrn: Deep complex convolution recurrent network for phase-aware speech
  enhancement,'' \emph{arXiv preprint arXiv:2008.00264}, 2020.

\bibitem{luo2020dual}
Y.~Luo, Z.~Chen, and T.~Yoshioka, ``Dual-path rnn: efficient long sequence
  modeling for time-domain single-channel speech separation,'' in \emph{ICASSP
  2020-2020 IEEE International Conference on Acoustics, Speech and Signal
  Processing (ICASSP)}.\hskip 1em plus 0.5em minus 0.4em\relax IEEE, 2020, pp.
  46--50.

\bibitem{li2021icassp}
A.~Li, W.~Liu, X.~Luo, C.~Zheng, and X.~Li, ``Icassp 2021 deep noise
  suppression challenge: Decoupling magnitude and phase optimization with a
  two-stage deep network,'' in \emph{ICASSP 2021-2021 IEEE International
  Conference on Acoustics, Speech and Signal Processing (ICASSP)}.\hskip 1em
  plus 0.5em minus 0.4em\relax IEEE, 2021, pp. 6628--6632.

\bibitem{8384449}
H.~Bu, J.~Du, X.~Na, B.~Wu, and H.~Zheng, ``Aishell-1: An open-source mandarin
  speech corpus and a speech recognition baseline,'' in \emph{2017 20th
  Conference of the Oriental Chapter of the International Coordinating
  Committee on Speech Databases and Speech I/O Systems and Assessment
  (O-COCOSDA)}, 2017, pp. 1--5.

\bibitem{du2018aishell}
J.~Du, X.~Na, X.~Liu, and H.~Bu, ``Aishell-2: Transforming mandarin asr
  research into industrial scale,'' \emph{arXiv preprint arXiv:1808.10583},
  2018.

\bibitem{brummer2013bosaris}
N.~Br{\"u}mmer and E.~De~Villiers, ``The bosaris toolkit: Theory, algorithms
  and code for surviving the new dcf,'' \emph{arXiv preprint arXiv:1304.2865},
  2013.

\bibitem{thienpondt2021idlab}
J.~Thienpondt, B.~Desplanques, and K.~Demuynck, ``The idlab voxsrc-20
  submission: Large margin fine-tuning and quality-aware score calibration in
  dnn based speaker verification,'' in \emph{ICASSP 2021-2021 IEEE
  International Conference on Acoustics, Speech and Signal Processing
  (ICASSP)}.\hskip 1em plus 0.5em minus 0.4em\relax IEEE, 2021, pp. 5814--5818.

\end{thebibliography}

\end{document}